# BYOD, Personal Area Networks (PANs) and IOT: Threats to Patients Privacy

Samara M. Ahmed, MD

Assistant Professor, College of Medicine

King Abdulaziz University, Jeddah

## Abstract

The passage of FISMA and HIPPA Acts have mandated various security controls that ensure the privacy of patients' data. Hospitals and health-care organizations are required by law to ensure that patients' data is stored and disseminated in a secure fashion. The advent of Bring Your Own Devices (BYOD), mobile devices, instant messaging (such as WhatsApp) and cloud technology however, have brought forth new challenges. The advent of Internet of Things (IOT) have complicated the matters further as organizations are not fully cognizant to the all facets of threats to data privacy. Physicians and health care practitioners need to be made aware of various new avenues of data storage and transmission that need to be secured and controlled. In this paper we look at various threats and challenges that IOT, Bring Your Own Device (BYOD) and Personal Area Networks (PANs) technologies pose to the patients' privacy data. We conclude the paper by providing the results of a survey that gauge the depth of understanding of healthcare professionals regarding the emerging threats to patients' privacy

## 1. Introduction

The networked nature of healthcare environment has allowed medical and healthcare practitioners to exchange patients' information across various platforms. The promise of web3.0 [22] brings a plethora of opportunities to glean useful information from historical data that is available to hospitals and healthcare organizations. Work done in [15] shows the potential to detecting mental health symptoms by scavenging through twitter data. Authors in [2] provide an overview of various efforts that have been carried out in this vein.

A smart healthcare environment has evolved where healthcare services are delivered seamlessly. Not only, the environment provides seamless access to information to healthcare professionals and the patients, but also provides a platform to help monitor patients remotely and at times also deliver medications automatically [16, 17]. The use of storage and network technologies is a prerequisite for such environment. However, such technologies put the data that is being stored at risks and in turn increases the risk of putting patients' privacy at risk. The Health Insurance and Portability Accountability Act (HIPAA) was passed in 1996 [23] that articulated various regulations that would help safeguard patients' privacy. The Federal Information Security Management Act of 2002 (FISMA) [24] was passed to augment the guidelines set out in HIPAA. Both the laws complimented each other in terms of putting various security controls in place.

The FISMA legislation was passed in 2002 and the dramatic change witnessed by the FISMA act witnessed many areas that were not covered. Beginning with the mobile devices and the control they provided, sensors were introduced in medical devices and even human body for exchange of sensitive patient data. These technologies pushed the bolstering of the FISMA initiative with Cyber Enhancement Act of 2014 [25]. The act further refined the security controls that the federal agencies must put in place. The Food and Drug Administration (FDA) published non-binding requirements for medical devices [26] and also supported the Mitre corporation effort that specifically focusses on making medical devices more secure [27].

Securing the medical devices is a small part of the security equation however. The study conducted in [9] provided results on lack of understanding on part of people on the working of internet and the various components that are involved and how majority of the people are unaware of the underlying risks in the new healthcare environment. The growth of Internet of Things (IoT) paradigm complicated the matters further as it allowed various mobile devices and sensors to monitor heterogeneous systems and humans in real-time. While practitioners can now practice preventive medicine via IoT platform and at times even deliver medication [13,14], little thought has been given to ensuring a secure exchange of such information. The underlying architecture for such enabling technologies further complicates the comprehension of the vast landscape that underscores a smart health care environment. This paper aims at addressing the following:

1. Describe a typical smart healthcare architecture and its components
2. Describe legislations and other efforts that put policies and procedures in place
3. Challenges provided by PAN and BYOD devices
4. Initial surveys

## 2. Typical Smart Healthcare Architecture

As mentioned in [9], one of the biggest challenges when dealing with patients' data and the underlying privacy is the lack of understanding as to how the information is transmitted and stored via a network. Figure 1 taken from [17] shows a typical smart healthcare environment.

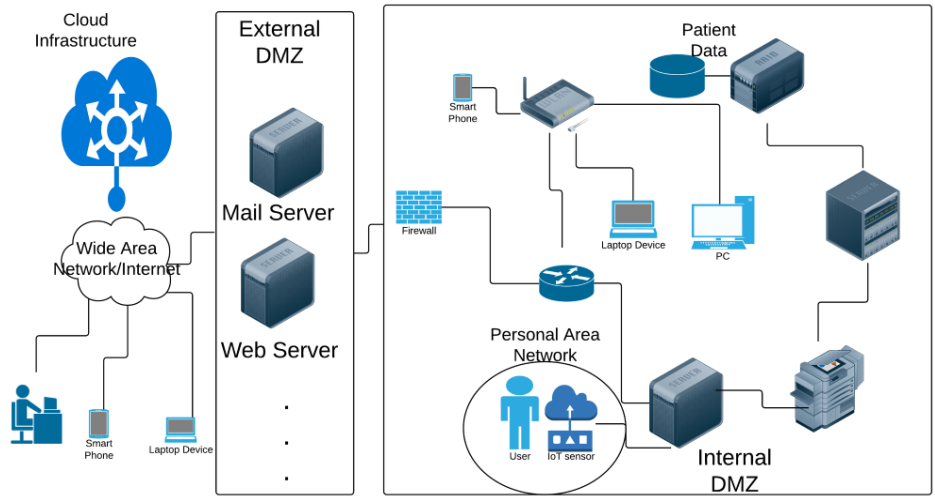

*Figure 1: A typical smart health-care architecture*

The study grouped the threats to patients' data under the following two categoris namely 1) Network and 2) Technology. For our work, we will focus on the Network layer only.

### 2.1. Network Layer

The Network layer can be further divided into the following three areas.

1. Local Area Network (LAN)
2. Wide Area Network (WAN)
3. Personal Area Network (PAN)

We briefly describe and explain the LAN and WAN networks as this topic has been discussed in detail in literature and traditional security solutions have focused on such technologies. We will expand upon the PAN aspect as it introduces more vulnerability to the patients' data

### 2.1.1. Local Area Network (LAN)

The LAN environment is the internal network of an organization and is typically hidden from outside the enterprise boundaries. Employees can connect to the corporate LAN within the organization or employ techniques such as Virtual Private Network when connecting over the Internet (encrypting the link that allows the user to access the digital assets remotely). A VPN is considered relatively safe as the organization sets it up per the best practices usually and is difficult to hack

### 2.1.2. Wide Area Network (WAN)

WAN has become synonymous with the word Internet. Globally people can access various digital resources and while some corporations might have setup private WANs, majority of the organizations utilize the global network and depend upon Internet Service Providers (ISPs) to allow their corporate users to reach and use the resources on their private local area networks.

## 2.1.3. Personal Area Network (PAN)

A much smaller network in size, A PAN is limited to within 10-20 meters of the person or device. These networks are characterized by a sensor and an actuator. The sensor's job is to notice changes in certain stimuli (e.g., heart rate of a person, temperature of a room etc.). The actuator is the physical device that would move once a trigger is activated. For example, the thermostat would cause the AC to run off once the temperature falls below a certain range and activate a trigger.

The presence of a PAN is usually dependent upon either a mobile device with the user or a device that is within the proximity of the patient. As pointed out by [1], there are primarily four standards that help in forming a PAN.

1. Bluetooth
2. IEEE 802.1.5.4 Low rate WPAN
3. IEEE 802.15.3 High rate WPAN
4. IEEE 802.15. 6 Body Area Networks

National Institute of Science and Technology (NIST) [28] summarizes the risks that the mobile devices can cause to the corporate networks in SP800-46. These include:

1. Lack of Physical Security Controls: This refers to the fact that the device used by the end user that can connect to the LAN and PAN over the Internet can easily be stolen and accessed by a malicious user
2. Unsecured Networks and Man in The Middle (MITM) attacks: The device used by the healthcare practitioners can easily be sniffed by someone on both the WAN and the PANs. The security level on the Internet cannot be controlled by the enterprise while the PANs are relatively new and can be compromised with a much more ease by the malicious user
3. Infected Devices: The mobile device used by the healthcare practitioner can be infected with malicious software. The device having access to the PANs allow the malicious user to access the PAN directly and in turn get access to the patient data.

Furthermore, most of the mobile devices are connected to the cloud which can possible have lax security controls in place this endangering the patient private data

## 3. Related Legislation

Various laws have been formulated by the United States Congress that mandate the privacy requirements for patients' data. While the jurisdiction of such laws is limited to the US, similar laws are being adopted globally. We briefly present the applicable laws below.

### 3.1. Applicable Laws

## 3.1.1. HIPAA

The HIPAA law [23] for the first time discussed how the privacy of patient needs to be preserved and the patient need to be informed of how his information is handled and disseminated. One of the corollaries of the HIPAA law was the effect it had on information stored in IT systems and the safe storage and transmission of the patients' information.

### 3.1.2. Federal Information Security Management Act of 2002

The FISMA act (also known as the E-Government Act of 2002) zeroed in on the electronic services provided by the government and established the governance structure to propagate the importance of information security and privacy [24]. Given the networked nature of the various IT systems in place and the associated security controls needed in place, FISMA delegated the assigned the task of developing various standards and policies to National Institute of Science and Technology (NIST). Such standards and policies are paramount to put information security controls in place.

### 3.1.3. Cyber Enhancement Act 2014

The Cyber Enhancement Act (CEA) of 2014 [25] augments the FISMA 2002 Act and emphasizes further areas of focus. The act addresses the physical aspect of information security and also briefly addresses cyber-physical systems (backbone of the Personal Area Networks) [1]. The CEA Act also recognizes the nascent nature of the recent standards and thus emphasizes further research in this area. This work is a step in this direction.

### 3.2. NIST Recommendations

NIST is a non-regulatory body under the US Department of Commerce and is tasked with developing policies and recommendations for information security controls. NIST publishes two set of documents namely Federal Information Processing Standards (FIPS) series and the Special Publications 800 (SP-800) series. The FIPS series is a list of mandatory standards for federal organizations while the SP-800 series recommends various controls that can be implemented. The following table taken from [1] summarizes the relevant documents from NIST [29,30].

| Title | Description |
|---|---|
| **FIPS-140** | Security Requirements for Cryptography Modules |
| **FIPS-199** | Standards for Security Categorization of Federal Information and Information Systems |
| **FIPS 200** | Minimum Security Requirements for Federal Information and Information Systems |
| **FIPS 201** | Personal Identity Verification for Employees and Contractors |
| **SP800-12** | An Introduction to Information Security |
| **SP800-18** | Guide for Developing Security Plans for Federal Information Systems |
| **SP800-30** | Guide for Conducting Risk Assessments |
| **SP800-32** | Introduction to Public Key Technology and the Federal PKI Infrastructure |
| **SP800-37** | Risk Management Framework for Information Systems and Organizations: A System Life Cycle Approach for Security and Privacy |
| **SP800-39** | Managing Information Security Risk: Organization, Mission, and Information System View |
| **SP800-46** | Guide to Enterprise Telework, Remote Access, and Bring Your Own Device (BYOD) Security |
| **SP800-53** | Security and Privacy Controls for Federal Information Systems and Organizations |
| **SP800-66** | An Introductory Resource Guide for Implementing the Health Insurance Portability and Accountability Act (HIPAA) Security Rule |
| **SP800-121** | Guide to Bluetooth Security |
| **SP800-183** | Networks of 'Things' |
| **SP800-187** | Guide to LTE Security |

## 4. Healthcare Professionals - Initial Survey Results

Given the above and similar to [9], we carried a brief survey of healthcare professionals and got the following results.

| | |
|---|---|
| Number of Participants | 50 |
| Awareness of Cloud Risks | 6% |
| Using WhatsApp Application to discuss patient info | 69% |
| Perpetual Bluetooth Device usage | 82% |

Looking at the above data, we gathered the following:
1. Majority of the healthcare professionals are using instant messaging technology to discuss patients' cases

2. Healthcare professionals are unaware of the risk posed by the Bluetooth technology
3. Healthcare professionals are unaware of the amount/type of data that is being stored in the cloud

Please note the following:
1. Before proceeding to get the ethics approval, we did an informal survey on survey monkey that was completely anonymous
2. The sample size was relatively small but enough to establish a pattern
3. The sample size was not specific to a particular institution but rather spread across many institutions
4. The sample size included a range of healthcare professionals and was not limited to physicians

# 5. Conclusion

The challenges to preserving the patients' data security and privacy are quite a few. However, given the explosion in the network and mobile technology has given rise to unprecedented challenges. In this paper, we have provided an overview of such technologies and more importantly we have set the platform for conducting a more through survey to gauge the healthcare professionals' insight into the threats of patients' privacy and the effects of HIPAA and other related laws on the usage of emerging technologies. We hope to expand on thos work and explore this topic further.

# 6. References

[1] Rajput, A., & Brahimi, T. (2019). Characterizing IOMT/Personal Area Networks Landscape. In M. Lytras et al. (Eds.) *Innovation in Health Informatics: a Smart Healthcare Primer*. Amsterdam, Netherlands: Elsevier. (earlier version available as *arXiv preprint arXiv:1902.00675)*.

[2] Rajput, A. E., & Ahmed, S. M. (2019). Big Data and Social/Medical Sciences: State of the Art and Future Trends. *arXiv preprint arXiv:1902.00705*.

[3] Huang, L. C., Chu, H. C., Lien, C. Y., Hsiao, C. H., & Kao, T. (2009). Privacy preservation and information security protection for patients' portable electronic health records. *Computers in Biology and Medicine*, *39*(9), 743-750.

[4] Haas, S., Wohlgemuth, S., Echizen, I., Sonehara, N., & Müller, G. (2011). Aspects of privacy for electronic health records. *International journal of medical informatics*, *80*(2), e26-e31.

[5] Fernández-Alemán, J. L., Señor, I. C., Lozoya, P. Á. O., & Toval, A. (2013). Security and privacy in electronic health records: A systematic literature review. *Journal of biomedical informatics*, *46*(3), 541-562.

[6] Anwar, M., Joshi, J., & Tan, J. (2015). Anytime, anywhere access to secure, privacy-aware healthcare services: Issues, approaches and challenges. *Health Policy and Technology*, *4*(4), 299-311.

[7] Xu, L., Jiang, C., Wang, J., Yuan, J., & Ren, Y. (2014). Information security in big data: privacy and data mining. IEEE Access, 2, 1149-1176.

[8] Yüksel, B., Küpçü, A., & Özkasap, Ö. (2017). Research issues for privacy and security of electronic health services. *Future Generation Computer Systems*, *68*, 1-13.

[9] Kang, R., Dabbish, L., Fruchter, N., & Kiesler, S. (2015, July) "My data just goes everywhere:" user mental models of the internet and implications for privacy and security. In *Symposium on Usable Privacy and Security (SOUPS)* (pp. 39-52). Berkeley, CA: USENIX Association.

[10] Huang, L. C., Chu, H. C., Lien, C. Y., Hsiao, C. H., & Kao, T. (2009). Privacy preservation and information security protection for patients' portable electronic health records. *Computers in Biology and Medicine*, *39*(9), 743-750.
[11] Arias, O., Wurm, J., Hoang, K., & Jin, Y. (2015). Privacy and security in internet of things and wearable devices. *IEEE Transactions on Multi-Scale Computing Systems*, *1*(2), 99-109.
[12] Farooq, M. U., Waseem, M., Khairi, A., & Mazhar, S. (2015). A critical analysis on the security concerns of internet of things (IoT). *International Journal of Computer Applications*, *111*(7).
[13] Hussain, A., Wenbi, R., da Silva, A. L., Nadher, M., & Mudhish, M. (2015). Health and emergency-care platform for the elderly and disabled people in the Smart City. *Journal of Systems and Software*, *110*, 253-263.
[14] Doukas, C., Metsis, V., Becker, E., Le, Z., Makedon, F., & Maglogiannis, I. (2011). Digital cities of the future: Extending@ home assistive technologies for the elderly and the disabled. *Telematics and Informatics*, *28*(3), 176-190.
[15] Rajput, A., & Ahmed, S. (2019). Making a Case for Social Media Corpus for Detecting Depression. *arXiv preprint arXiv:1902.00702*.
[16] Rajput, A. (2019). Natural Language Processing, Sentiment Analysis and Clinical Analytics. In M. Lytras et al. (Eds.) *Innovation in Health Informatics: a Smart Healthcare Primer*. Amsterdam, Netherlands: Elsevier. (available as *arXiv preprint arXiv:1902.00679*)
[17] Ahmed, S. (2019) Threats to Patients Privacy in Smart Healthcare Environment. In M. Lytras et al. (Eds.) *Innovation in Health Informatics: a Smart Healthcare Primer*. Amsterdam, Netherlands: Elsevier.
[18] Sadan, B. (2001). Patient data confidentiality and patient rights. *International journal of medical informatics*, *62*(1), 41-49.
[19] Sahi, A., Lai, D., & Li, Y. (2016). Security and privacy preserving approaches in the eHealth clouds with disaster recovery plan. *Computers in biology and medicine*, *78*, 1-8.
[20] Wu, R. (2012). *Secure sharing of electronic medical records in cloud computing*. Arizona State University.
[21] Sweeney, L. (2002). k-anonymity: A model for protecting privacy. *International Journal of Uncertainty, Fuzziness and Knowledge-Based Systems*, *10*(05), 557-570.
[22] Lassila, O., & Hendler, J. (2007). Embracing" Web 3.0". *IEEE Internet Computing*, *11*(3).
[23] https://www.gpo.gov/fdsys/pkg/BILLS-104s1028is/pdf/BILLS-104s1028is.pdf
[24] https://csrc.nist.gov/topics/laws-and-regulations/laws/fisma
[25] https://www.gpo.gov/fdsys/pkg/PLAW-113publ283/pdf/PLAW-113publ283.pdf
[26] https://www.fda.gov/downloads/MedicalDevices/DeviceRegulationandGuidance/GuidanceDocuments/UCM623529.pdf
[27] https://www.mitre.org/publications/technical-papers/medical-device-cybersecurity-regional-incident-preparedness-and
[28] www.nist.gov
[29] https://csrc.nist.gov/publications/fips
[30] https://csrc.nist.gov/publications/sp800